\documentclass[a4paper]{jpconf}
\usepackage{graphicx}
\usepackage{leftidx}
\usepackage{cite}
\usepackage{amsmath,amssymb,amsopn,bm,bbm,slashed}

\newcommand{\dd}{\ensuremath{\mathrm{d}}}
\newcommand{\MeV}{\ensuremath{\mathrm{MeV}}}

\newcommand{\SU}{\ensuremath{\mathrm{SU}}}
\newcommand{\ii}{\ensuremath{\mathrm{i}}}
\newcommand{\fslash}{\slashed}
\newcommand{\erw}[1]{\ensuremath {\left \langle {#1} \right \rangle}}

\begin{document} 
\title{Monte-Carlo approach to particle-field interactions and the
  kinetics of the chiral phase transition}

\author{Carsten Greiner$^1$, Christian Wesp$^1$, Hendrik van
  Hees$^{1,2}$, Alex Meistrenko$^1$}
\address{$^1$Institut f{\"u}r theoretische Physik, Goethe-Universit{\"a}t
  Frankfurt am Main, Max-von-Laue-Stra{\ss}e 1, D-60438 Frankfurt, Germany1}
\address{$^2$Frankfurt Institute for Advanced Studies, Ruth-Moufang-Stra{\ss}e, D-60438 Frankfurt,
  Germany}

\ead{Carsten.Greiner@th.physik.uni-frankfurt.de}

\begin{abstract}
  The kinetics of the chiral phase transition is studied within a linear
  quark-meson-$\sigma$ model, using a Monte-Carlo approach to
  semiclassical particle-field dynamics. The meson fields are described
  on the mean-field level and quarks and antiquarks as ensembles of test
  particles. Collisions between quarks and antiquarks as well as the
  $q\overline{q}$ annihilation to $\sigma$ mesons and the decay of
  $\sigma$ mesons is treated, using the corresponding transition-matrix
  elements from the underlying quantum field theory, obeying strictly
  the rule of detailed balance and energy-momentum conservation. The
  approach allows to study fluctuations without making ad hoc
  assumptions concerning the statistical nature of the random process as
  necessary in Langevin-Fokker-Planck frameworks.
\end{abstract}

\section{Introduction}

One of the motivations for the study of ultrarelativistic heavy-ion
collisions is to gain a detailed understanding of the phase diagram of
strongly interacting matter\cite{Friman:2011zz}. At the largest energies
as achieved at the Large Hadron Collider (LHC) and the Relativistic
Heavy Ion Collider (RHIC) a hot and dense fireball is formed which can
be described to a surprising accuracy as a nearly perfect fluid of
strongly coupled quarks and gluons (QGP) undergoing a transition to a
hot hadron-resonance gas. In these situations, where the net-baryon
density or the baryon-chemical potential are small, lattice-QCD (lQCD)
calculations indicate a crossover transition from confined to deconfined
matter as well as from a phase where chiral symmetry is spontaneously
broken to one where it is restored at a (common) transition temperature
$T_{\mathrm{c}} \simeq 160 \; \MeV$\cite{Philipsen:2012nu}.

At lower collision energies, as studied in the RHIC beam-energy scan
(BES) program, at the CERN SPS, and the future FAIR and NICA
experiments, the produced medium starts at lower temperatures and larger
net-baryon densities. Since in this situation the application of lQCD is
challenging due to the ``sign problem'' at finite $\mu_{\text{B}}$, one
relies on effective chiral models, which predict the existence of a
first-order chiral-phase-transition line ending in a critical point of a
second-order phase transition.

For theory the challenge is to provide possible observables for this
phase structure in heavy-ion collisions, like the (``grand-canonical'')
fluctuations of conserved charges like net-baryon number or electric
charge. Not only the question, how to effectively model the phase
transition (e.g., with the Nambu-Jona-Lasinio (NJL) or the (linear)
$\sigma$ model with extensions taking into account gluonic degrees of
freedom implementing Polyakov loops) arises but also, which of the
features of the phase structure predicted for such models applying
thermal quantum field theory (describing a medium in thermal and
chemical equilibrium) like (critical) fluctuations of conserved charges
survive for a rapidly expanding and cooling fireball as created in
heavy-ion collisions.

To address the latter question, one relies on transport simulations to
describe the off-equilibrium dynamics of the fireball. One approach is
the use of ideal or viscous hydrodynamics to describe the bulk evolution
of the fireball (assuming a state close to local thermal equilibrium),
which successfully describes key phenomena of heavy-ion collisions, and
adding the fluctuations by hand in a Langevin
approach\cite{Nahrgang:2011mg,Nahrgang:2011mv,Herold:2013bi}. On the
other hand this implies that the statistics of the random process has to
be put in as an ad hoc assumption. Usually a Gaussian Markovian (``white
noise'') is assumed, but the simulation of non-Markovian (``colored
noise'') processes is feasible in principle \cite{Schmidt:2014zpa}.

On the other hand, one would like to study the implication of different
transition scenarios (cross-over, 1$^{\text{st}}$ order, 2$^{\text{nd}}$
order) as realized in the various quantum-field theoretical models on
the nature of the hopefully observable fluctuations
\cite{Stephanov:1999zu,Schaefer:2007pw,Skokov:2010uh,Skokov:2010wb,BraunMunzinger:2011ta,Schaefer:2011pn,Morita:2012kt}
on the statistics of the probably observable fluctuations.

In this work we present a novel Monte-Carlo approach to address this
challenging problem using the most simple quark-meson linear $\sigma$
model \cite{Wesp:2014xpa}. The meson fields are treated on the
mean-field level and the quarks and antiquarks are realized in terms of
a test-particle ensemble. Here the challenge is to implement
``discrete'' local interaction processes like elastic collisions and
reactions like the $q\overline{q}$ annihilation to a $\sigma$ meson and
the decay of $\sigma$ mesons to a $q \overline{q}$ pair admitting not
only kinetic but also chemical equilibration starting from an
off-equilibrium situation, using the transition-probability matrix
elements of the underlying quantum field theory.

\section{Linear quark-meson $\sigma$ model}

To investigate the feasibility of a kinetic description of the
off-equilibrium dynamics of the chiral phase transition the most simple
two-flavor chiral model, based on the chiral group
$\SU(2)_{\text{L}} \times \SU(2)_{\text{R}}$ is considered, using a
flavor doublet of Dirac fields $\psi$, describing $u$ and $d$ quarks and
antiquarks and a four-dimensional real-valued set of scalar fields
$(\sigma,\vec{\pi})$ transforming under the SO(4) representation of the
chiral group \cite{GellMann:1960np}. The Lagrangian reads
\begin{equation}
\label{1}
\mathcal{L}=\overline{\psi} [\ii \fslash{\partial}-g(\sigma + \ii
\gamma_5 \vec{p} \cdot \vec{\tau})] \psi + \frac{1}{2} (\partial_{\mu}
\sigma \partial^{\mu} \sigma + \partial_{\mu} \vec{\pi}
\cdot \partial^{\mu} \vec{\pi}) - U(\sigma,\vec{\pi})
\end{equation}
with the meson potential
\begin{equation}
\label{2}
U(\sigma,\vec{\pi}) = \frac{\lambda^2}{4} (\sigma^2+\vec{\pi}^2-\nu^2)^2
- f_{\pi} m_{\pi}^2 \sigma-U_0,
\end{equation}
where $g \in [3.3,5.5]$ denotes the Yukawa coupling between quarks and
mesons, $\lambda^2=20$ the meson coupling constant (corresponding to a
$\sigma$ mass of $m_{\sigma} \simeq 600 \; \MeV$), $f_{\pi}$ the
pion-decay constant, and $\nu^2=f_{\pi}^2-m_{\pi}^2/\lambda^2$. The
potential (\ref{2}) contains the explicit breaking of the chiral
symmetry due to the finite current quark masses resulting in a non-zero
pion mass, $m_{\pi} \simeq 138 \; \MeV$, of the pseudo-Goldstone modes
$\vec{\pi}$. The constituent quark masses are given by
$m_q^2 =g^2 \sigma_0^2$.

\begin{figure}
\begin{minipage}{0.49 \linewidth}
\includegraphics[width=\textwidth]{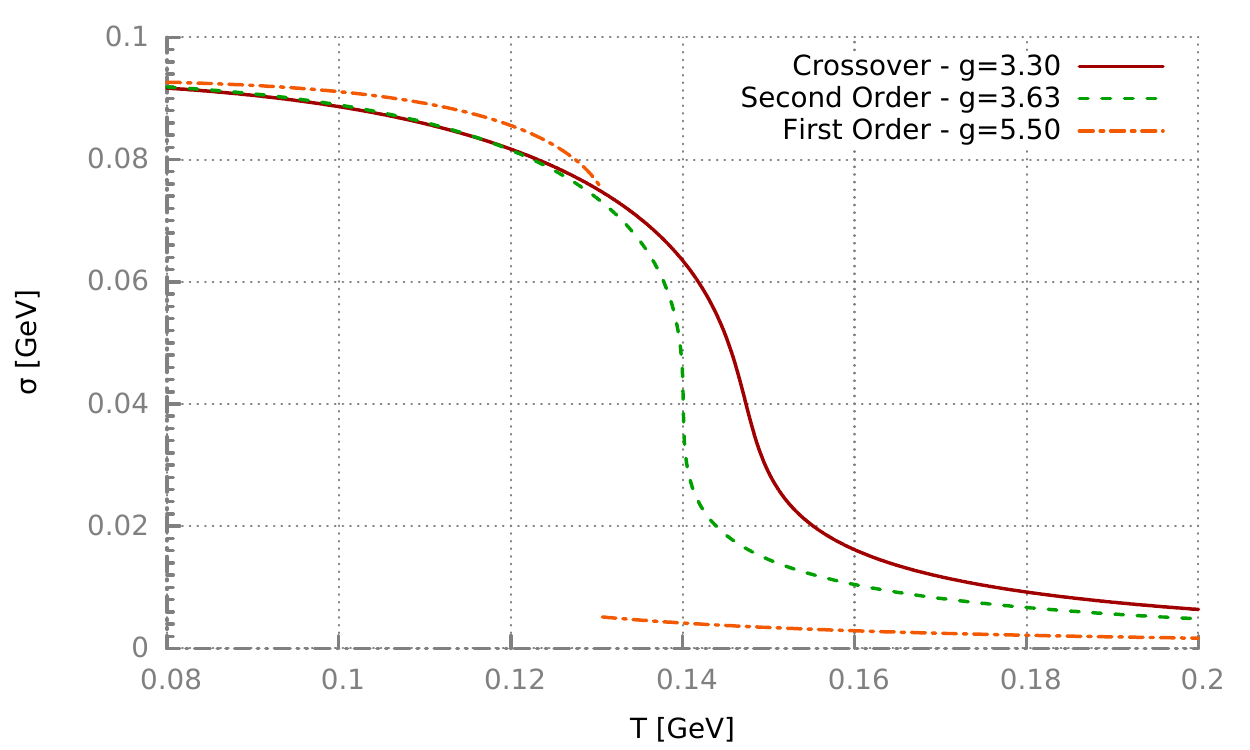}
\end{minipage}\hfill
\begin{minipage}{0.49 \linewidth}
\includegraphics[width=\textwidth]{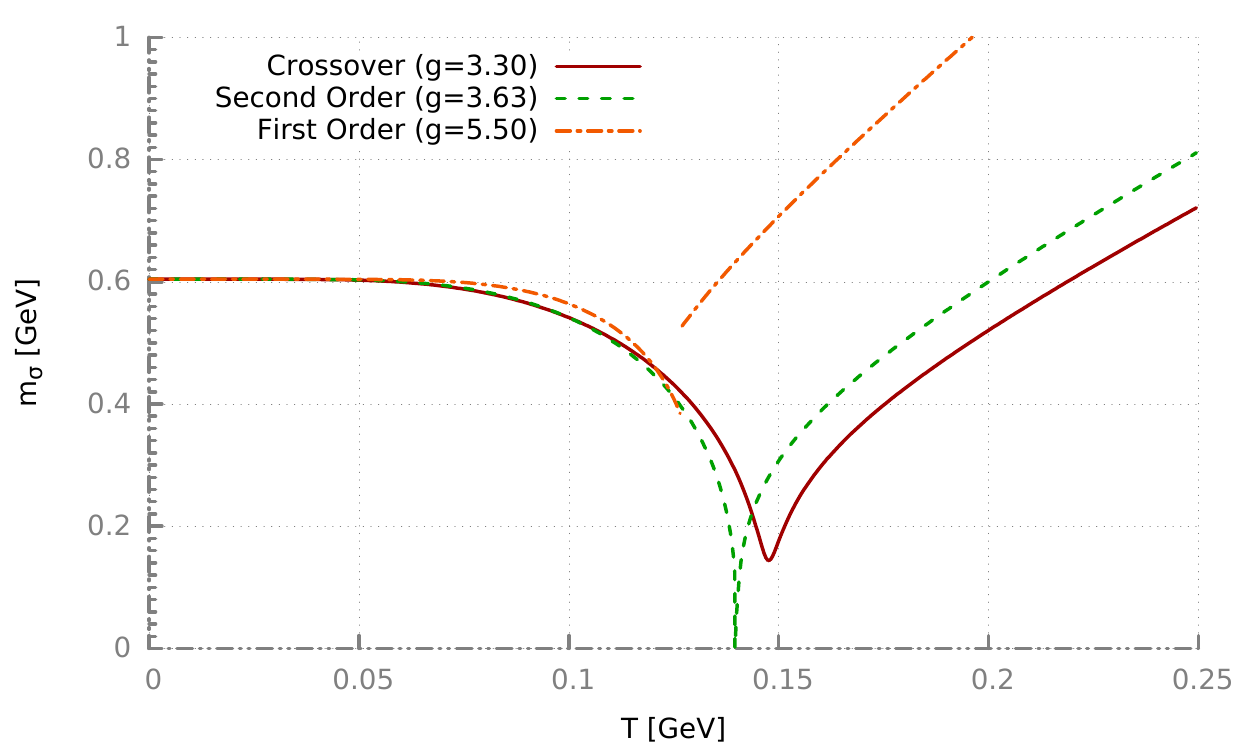}
\end{minipage}
\caption{\label{fig.1} The phase diagram for the linear $\sigma$ model
  in mean-field approximation at $\mu_{\mathrm{B}}=0$,
  $\erw{\vec{\pi}}=0$. Left: the order parameter $\erw{\sigma}$ and the
  effective $\sigma$ mass
  $m_{\sigma}=\partial^2 \Omega/\partial \sigma^2$.  }
\end{figure}

The grand-canonical potential in mean-field approximation reads
\begin{equation}
\label{3}
\Omega(T,\mu) = U(\sigma,\vec{\pi}) + \Omega_{\overline{\psi} \psi}
\end{equation}
with
\begin{equation}
\label{4}
\Omega_{\overline{\psi} \psi} = -d_n \int \frac{\dd^3 \vec{p}}{(2
  \pi)^3} \left [E + T \ln (1+\exp(-\beta(E-\mu)))+T \ln
  (1+\exp(-\beta(E+\mu))) \right ],
\end{equation}
where $E=\vec{p}^2+g^2 (\sigma^2+\vec{\pi}^2)$ and the quark-degeneracy
factor $d_n=2N_f N_c=12$. The mean fields have to be evaluated
self-consistently from the equilibrium condition
\begin{equation}
\label{5}
\frac{\partial \Omega}{\partial \sigma}=\frac{\partial \Omega}{\partial
  \vec{\pi}} = 0.
\end{equation}
In the following we restrict ourselves to vanishing pion mean
fields. A nice feature of this model is that by varying the Yukawa
coupling $g$, one finds different kinds of phase transition as
illustrated in Fig. \ref{fig.1}.

\section{Semiclassical particle-field dynamics}

The challenge in applying the above model to an off-equilibrium
dynamical simulation of a system of particles (here quarks and
antiquarks) and mean fields (representing the mesons) is that in order
to reproduce the equilibrium-phase structure as depicted in \ref{fig.1}
as the stationary limit, one has to ensure that both kinetic and
chemical equilibration is possible through the introduction of the
appropriate elastic collision terms for $qq$ and $q \overline{q}$
scattering as well as quark-number changing processes such as
$q \overline{q} \leftrightarrow \sigma$. In a full kinetic approach this
is achieved by a set of coupled Boltzmann-Vlasov equations, which read
in our case schematically (again restricting ourselves to the case of
vanishing pion-mean fields)
\begin{alignat}{2}
\label{6}
&\Box \sigma + \lambda(\sigma^2-\nu^2) \sigma - f_{\pi} m_{\pi}^2 + g
\erw{\overline{\psi} \psi} = I(\sigma \leftrightarrow \overline{q} q),
\\
\label{7}
&\left [\partial_t + \frac{p}{E_q} \cdot \vec{\nabla}_{\vec{x}} -
\vec{\nabla}_{\vec{x}} E_{\psi}(t,\vec{x},\vec{p}) \cdot
\vec{\nabla}_{\vec{p}}) \right ] f_{q}(t,\vec{x},\vec{p} ) = C(\psi \psi
\rightarrow \psi \psi,\sigma \leftrightarrow \overline{q} q).
\end{alignat}
Here, $I$ and $C$ denote collision integrals contributing to the
meson-mean-field and quark-phase-space distribution functions
respectively.

In the following a novel scheme to Monte-Carlo simulate such a system of
kinetic equations is defined, where one describes the mesons solely with
a mean field and the quarks and antiquarks in terms of test
particles. While the elastic-collision term is realized in a
straight-forward way using the corresponding cross section from the
underlying linear $\sigma$ model, one has to find a way to realize the
interactions $q \overline{q} \leftrightarrow \sigma$ in such a scheme,
while still fulfilling energy-momentum conservation and the principle of
detailed balance, which are the fundamental principles constraining the
off-equilibrium dynamics and ensuring the proper (Maxwell-Boltzmann)
equilibrium limit.

In our recently developed model (Dynamical Simulation of a Linear Sigma
Model, DSLAM) this challenge is solved as follows: In order to properly
simulate the collision terms on the right-hand sides of Eqs.\ (\ref{6})
and (\ref{7}) we define a space-time grid. In each time step at each
spatial cell the cross section for the annihilation process
$q \overline{q} \rightarrow \sigma$ is used to stochastically determine
an energy-momentum transfer from the initial $q \overline{q}$ pair,
located in the cell. This energy-momentum change is transferred to the
$\sigma$ mean field in terms of an appropriate relativistic Gaussian
wave packet $\delta \sigma(t,\vec{x})$ (which simulates the gain term in
$I$), and the $q \overline{q}$ pair is taken out of the test-particle
ensemble (which simulates the corresponding loss term in $C$).
\begin{figure}[t]
\begin{minipage}{0.48\linewidth}
\includegraphics[width=\textwidth]{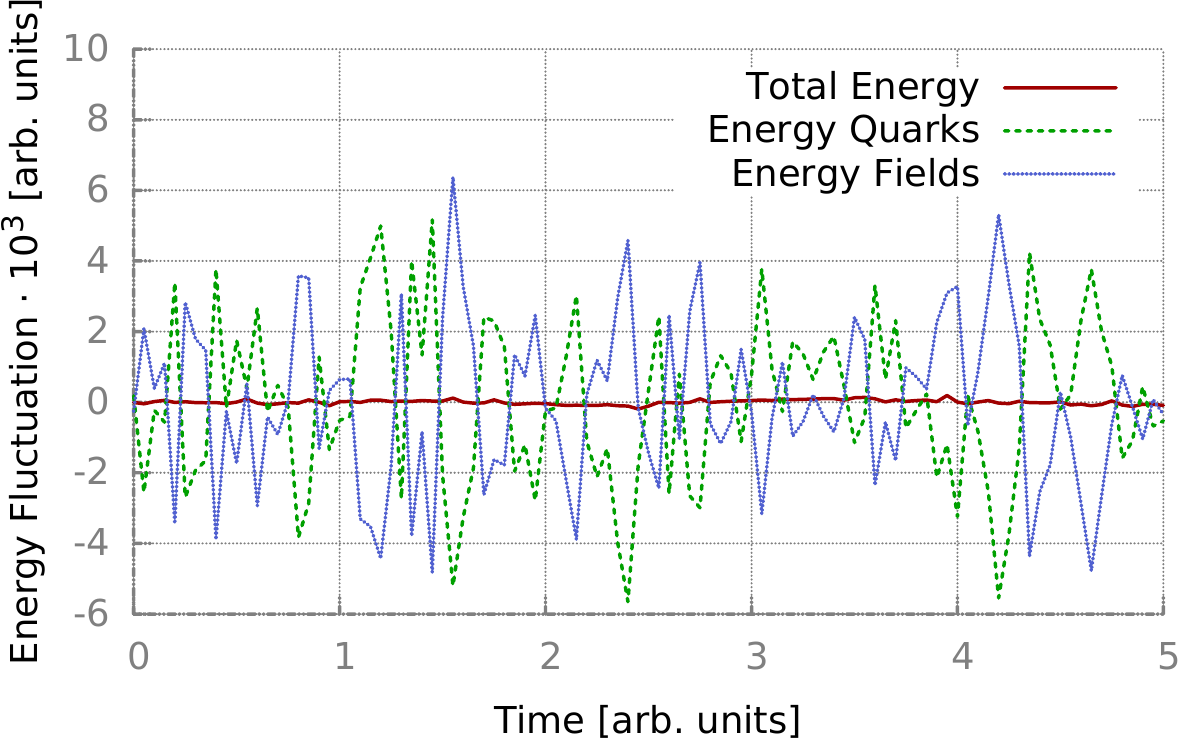}
\caption{\label{fig.2} While the energies of the quarks and the meson
  field in the thermal-box simulation show anti-correlated thermal
  fluctuations of the order $\Delta E/E \sim 10^{-3}$ for the quarks and
  $\Delta E/E \sim 10^{-2}$ for the field the numerical fluctuations of
  the total energy amount to only $\Delta E/e \lesssim 5 \cdot 10^{-5}$.
}
\end{minipage}\hfill
\begin{minipage}{0.48\linewidth}
\includegraphics[width=\textwidth]{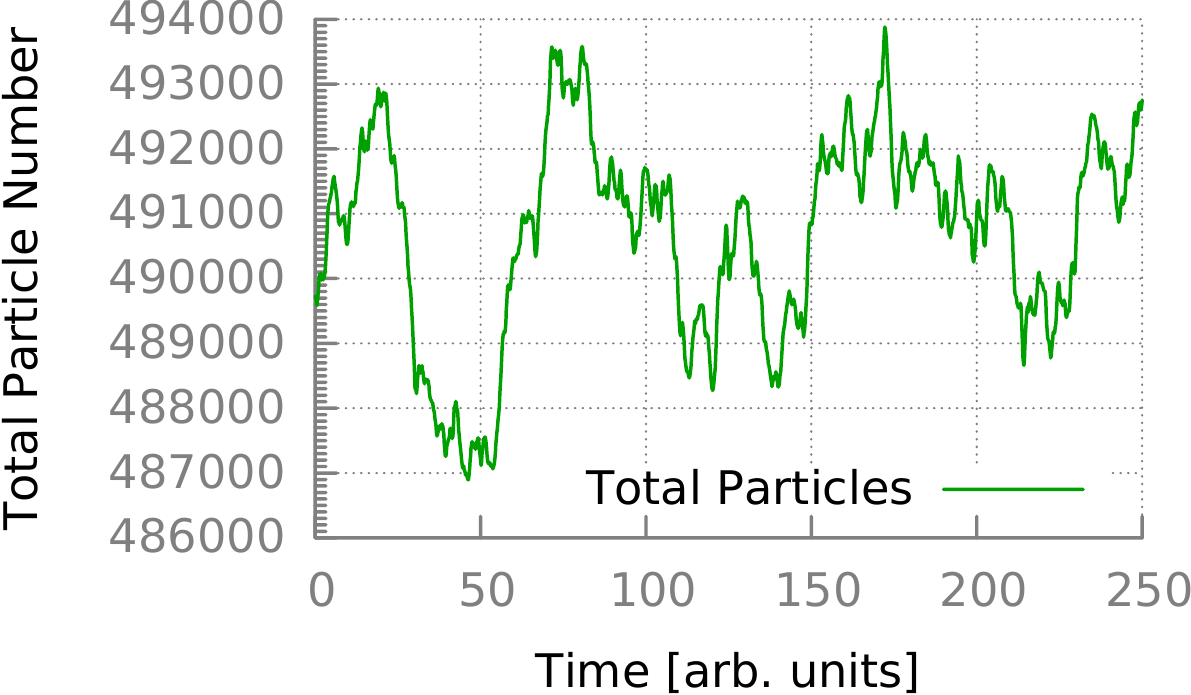}
\caption{\label{fig.3} The total quark number in the thermal-box
  simulation shows fluctuations due to the dynamical pair-creation and
  annihilation processes, leading to an exchange of energy between the
  quarks and the mean field as shown in the left Fig.\ \ref{fig.2}.
}
\end{minipage}
\end{figure}
To simulate also the appropriate decay process,
$\sigma \rightarrow \overline{q} q$, we have to ``particlize'' the mean
field locally in each spatial cell. This is done in the spirit of a
course-graining procedure: First, the total energy-momentum content of
the $\sigma$ field within the cell in terms of the corresponding
$\sigma$-field energy-momentum tensor is determined. Then one assumes a
local thermal equilibrium phase-space distribution, equivalent to this
energy-momentum tensor. In order to fulfill detailed balance, the
temperature has to be the same as that for the corresponding procedure
for the quarks and antiquarks. The temperature is related to the
mean-field value which depends on the scalar quark-antiquark density. In
this way a temperature can be determined. It is important to note that
it is defined in the local rest frame of the heat bath and thus the
$\sigma$-phase-space distribution is given by a Maxwell-J{\"u}ttner
distribution $f_{\sigma} \propto \exp[-p_{\mu} u^{\mu}(t,\vec{x})/T]$,
where $u^{\mu}$ is the four-velocity, given by the total
field-four-momentum in the spacial cell under consideration,
$u^{\mu}=p^{\mu}/E$. Now in each time-step within each spatial cell one
can choose an ensemble of $\sigma$ particles according to this local
Maxwell-J{\"u}ttner distribution and using the corresponding
$q \overline{q} \rightarrow \sigma$ decay rate to determine the gain
term to $C$. The loss term for the mean-field equation in the collision
term $I$ is again achieved by taking the appropriate amount of energy
and momentum out of the mean field in terms of a Gaussian wave packet.

In summary we have achieved a scheme which enables us to simulate the
set of Boltzmann-Vlasov Eqs. (\ref{6}) and (\ref{7}) using test
particles for the quarks and antiquarks and restricting the description
of the mesons strictly to the mean-field level. The scheme by
construction fulfills energy-momentum conservation through the Gaussian
wave-packet description for the exchange of energy and momentum between
the mean field and the test particles. At the same time also the
principle of detailed balance is fulfilled, using the coarse-graining
approach to locally map the field-energy-momentum distribution to a
local-equilibrium Maxwell-J{\"u}ttner distribution to reinterpret the
mean field as a meson phase-space distribution and using the
leading-order transition rates (cross sections) of the underlying QFT
linear $\sigma$ model fulfilling the detailed-balance principle.

\section{Proof of principle: ``Box'' calculations}

As a plausibility check of the simulation method the stability of an
equilibrium situation has been tested in a finite cubic box with
periodic boundary conditions\cite{Wesp:2014xpa}. The stability of the
energy conservation is demonstrated in Fig.\ \ref{fig.2}. While the
kinetic energy of the quarks and the energy of the mean field show
anti-correlated thermal fluctuations the total energy stays stable
within the numerical accuracy of the simulation.

The annihilation and creation processes of quark-antiquark pairs lead to
thermal fluctuations in the total particle number, as shown in Fig.\
\ref{fig.3}.
\begin{figure}[t]
\begin{minipage}{0.49\linewidth}
\includegraphics[width=\textwidth]{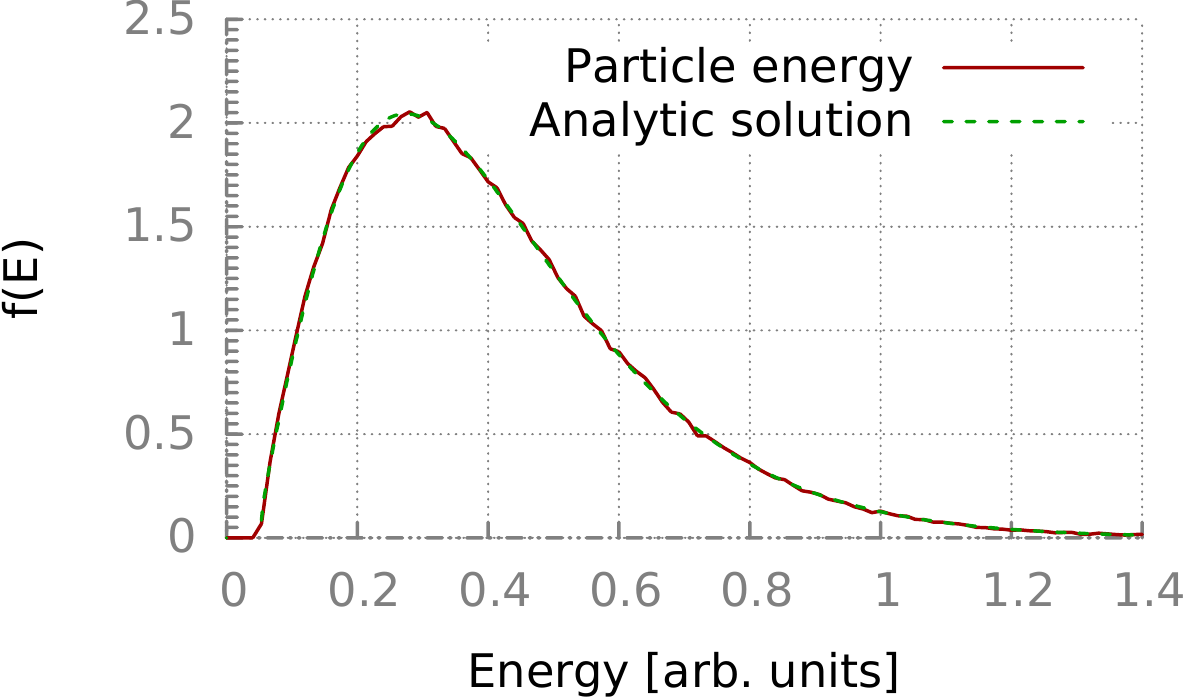}
\caption{\label{fig.4} The distribution of the quarks' kinetic energy in
  a thermal-box calculation show an excellent agreement with the expected
  Maxwell-Boltzmann distribution. 
}
\end{minipage}\hfill
\begin{minipage}{0.49\linewidth}
\includegraphics[width=\textwidth]{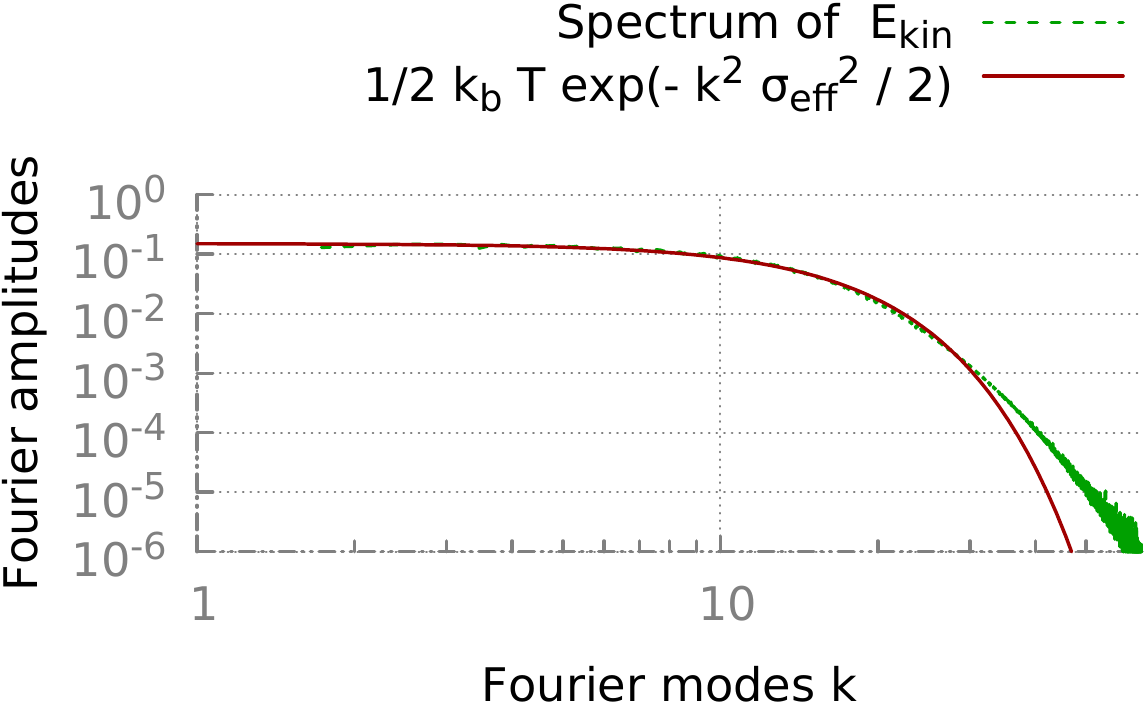}
\caption{\label{fig.5} The distribution of the kinetic field energy to
  its Fourier modes. The classical ``UV catastrophe'', i.e., the
  equipartition theorem associating an average energy of
  $k_{\text{B}} T/2$ for each Fourier mode, is avoided by the use of a
  finite finite width $\sigma_{\text{eff}}$ of the Gaussian wave packets
  for the transfer of energy and momentum between particles and field.
}
\end{minipage}
\end{figure}

The energy distribution of the quarks shows the expected
Maxwell-Boltzmann distribution at the expected temperature, as
demonstrated in Fig. \ref{fig.4}. Also the spectral analysis of the mean
field shows that the thermal equipartition theorem for the kinetic field
energy is fulfilled at small wave numbers (long wavelengths), i.e., each
mode contains an average energy of $k_{\text{B}} T/2$. On the other hand
the ``UV catastrophe'' must be avoided due to the finite total-energy
content within a finite box. Indeed, the energy-momentum transfer
between particles and the field due to the pair-annihilation and
-creation processes is not strictly ``local'' but occur within a finite
volume whose scale is fixed by the finite width, $\sigma$, of the
Gaussian wave packets used as field increments to keep care of the
correct energy-momentum transfer in each process. Thus the large wave
numbers (short wave lengths) are effectively cut off at a scale
$k_{\text{cutoff}}=\sqrt{2}/\sigma$, as shown by plotting the
corresponding Gaussian on top of the $k$-distribution of the kinetic
field energy (Fig.\ \ref{fig.5}).
\begin{figure}[t]
\begin{minipage}{0.4 \linewidth}
\includegraphics[width=0.49 \textwidth]{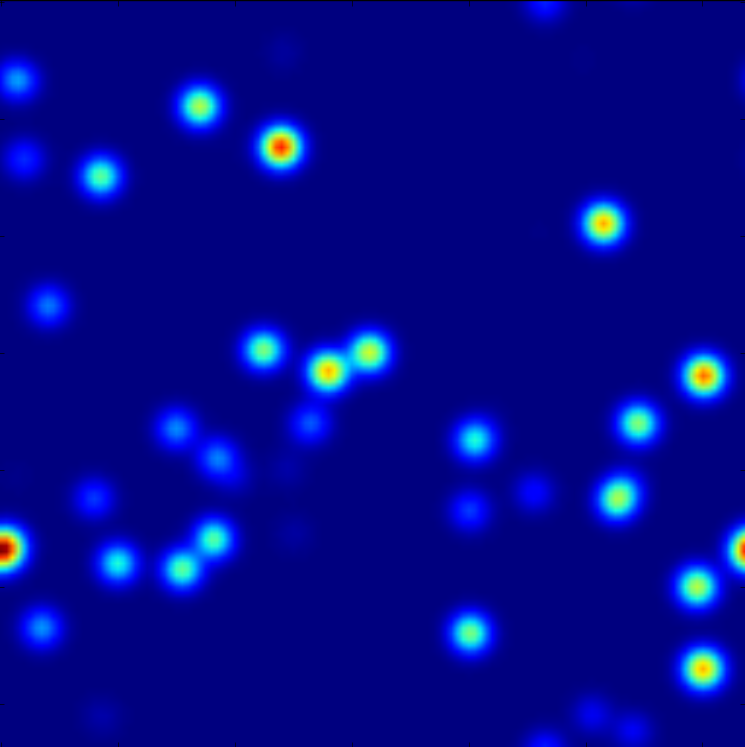}
\includegraphics[width=0.49 \linewidth]{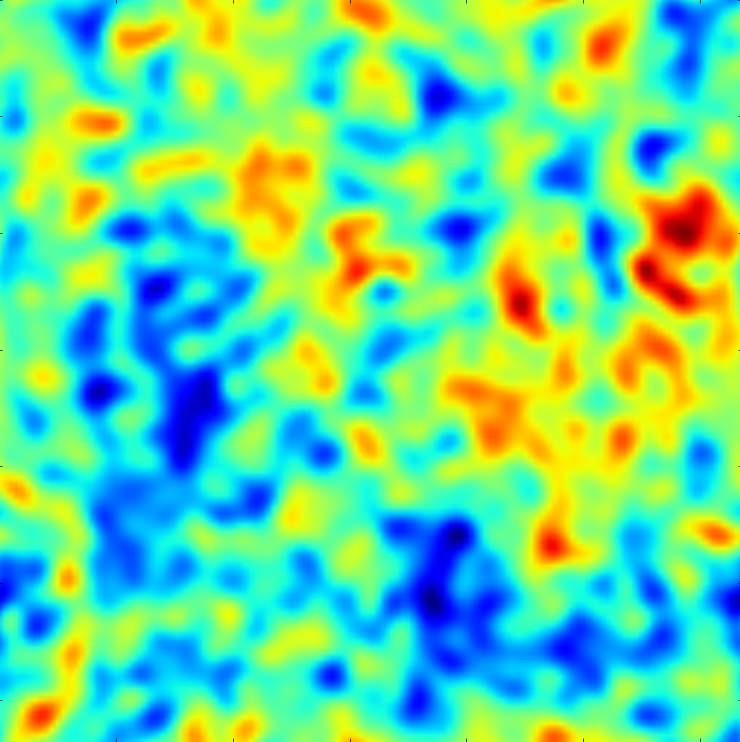}
\caption{\label{fig.6} Starting with a uniform mean field, a short time
  after the start of the simulation the field is disturbed by some
  Gaussian wave packets due to quark-pair annihilation processes (left
  panel). In the long-time limit the full equilibrium-thermal-field
  fluctuations have developed.}
\end{minipage}\hfill
\begin{minipage}{0.48 \linewidth}
\includegraphics[width= \textwidth]{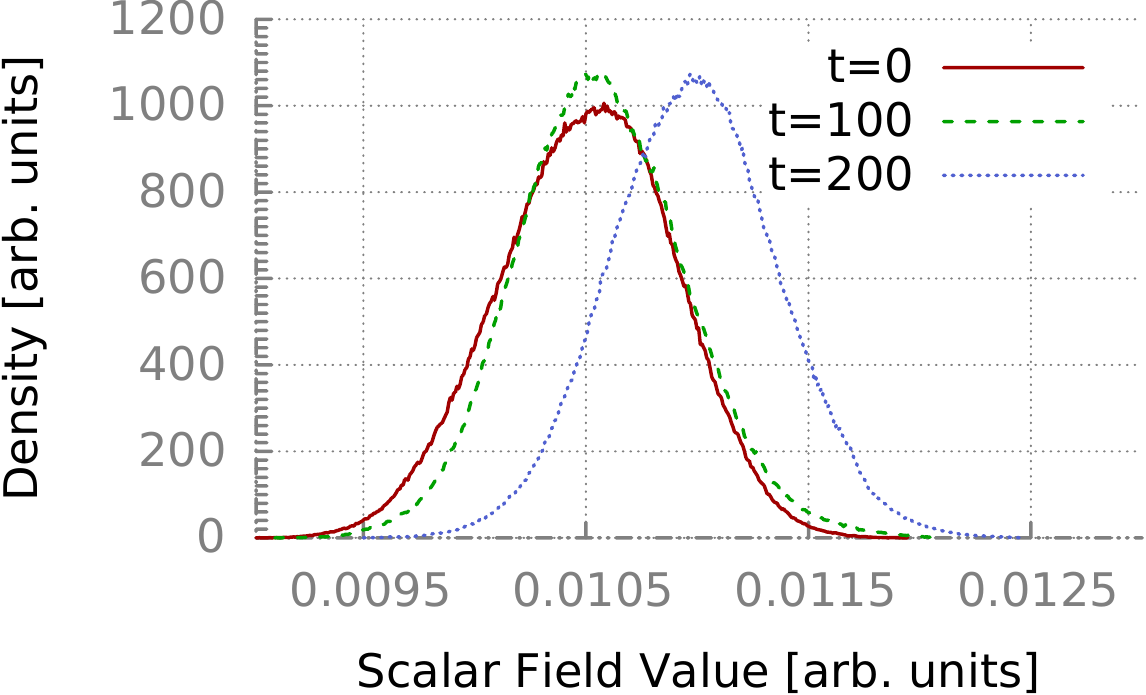}
\caption{\label{fig.7} The field fluctuations are Gaussian distributed
  around an average value which can slowly drift with time due to the
  dynamical fluctuations of the field momentum.  
}
\end{minipage}
\end{figure}

The dynamical behavior of the field in our thermal-box simulation is
illustrated in Fig.\ \ref{fig.6}, showing the field distribution within
the $xy$ plane of the simulation. Starting the simulation with a uniform
mean-field equilibrium value, after a short time some local blob-like
disturbances have developed (left panel) due to quark-pair-annihilation
processes, leading to the propagation of Gaussian wave packets on top of
the still quite uniform average field value, as implemented by our
concept of describing the exchange between field and particles in these
processes. The right panel shows the fully developed
equilibrium-thermal-field fluctuations in the long-time limit of the
simulation. In this limit the field values are Gaussian distributed
around a mean value (cf.\ Fig.\ \ref{fig.7}), which is expected due to
many random energy-momentum transfers between the field and
particles. Thereby the average field value can slowly drift with time
due to the thermal fluctuations of the field energy and momentum.

\section{Conclusions and outlook}

In this talk it was demonstrated that the dynamical description of the
chiral phase transition in a simple quark-meson model is feasible with a
novel Monte-Carlo-simulation technique for a corresponding coupled
Boltzmann-transport equation for the meson mean-field and the quark and
anti-quark phase-space distribution functions, implementing both elastic
quark and anti-quark scattering as well as quark-antiquark-pair creation
and annihilation processes, enabling both kinetic and chemical
equilibration between particles and the mean field (mesons).

The simulation is set up in a way that both energy-momentum conservation
and the principle of detailed balance are precisely realized (within the
limits of achievable numerical accuracy). While the elastic
quark-scattering processes are simulated with a straight-forward
test-particle realization, the particle-field kinetics has demanded the
development of a novel scheme.

The $q\overline{q} \rightarrow \sigma$ annihilation process is evaluated
by the Monte-Carlo sampling according to the corresponding transition
matrix elements from the underlying quantum-field theoretical
interpretation of the linear $\sigma$ model. The corresponding energy
and momentum are precisely transferred to the $\sigma$-mean field in
terms of an appropriate disturbance in form of a relativistic Gaussian
wave packet. To obey the principle of detailed balance, also the inverse
$\sigma \rightarrow q \overline{q}$ decay has to be simulated. To that
purpose a coarse-graining procedure based on the local energy-momentum
content of the field has been used for a ``particlization'' of the mean
field in terms of a local Boltzmann-J{\"u}ttner equilibrium
distribution, which in turn enables a Monte-Carlo sampling of the
$\sigma$-meson decay according to the decay rate from the quantum-field
theory picture of the model, which automatically implements detailed
balance.

This scheme has now been applied to off-equilibrium situations as in a
``thermal quench'' in a box where the particles and fields are
initialized with different temperatures and it was demonstrated that the
distribution can describe a possible phase transition from the initial
state to a late-time (equilibrium) state.

Last but not least also the case of ``expanding fireballs'', mimicking
the situation of the medium created in heavy-ion collisions, is under
study. 

\section*{Acknowledgments}

This work was partially supported by the Bundesministerium f{\"u}r
Bildung und Forschung (BMBF F{\"o}rderkennzeichen 05P12RFFTS) and by the
Helmholtz International Center for FAIR (HIC for FAIR) within the
framework of the LOEWE program (Landesoffensive zur Entwicklung
Wissenschaftlich-{\"O}konomischer Exzellenz) launched by the State of
Hesse.  C.\ W.\ and A.\ M.\ acknowledge support by the Helmholtz
Graduate School for Hadron and Ion Research (HGS-HIRe), and the
Helmholtz Research School for Quark Matter Studies in Heavy Ion
Collisions (HQM). Numerical computations have been performed at the
Center for Scientific Computing (CSC).  H.\ v.\ H.\ has been supported
by the Deutsche Forschungsgemeinschaft (DFG) under grant number GR
1536/8-1.  C.W. has been supported by BMBF under grant number 0512RFFTS.

\section*{References}

\begin{flushleft}

\providecommand{\newblock}{}

\end{flushleft}

\end{document}